\documentclass[twocolumn]{aastex63}
\usepackage{amsmath}
\usepackage{natbib}
\usepackage{savesym}
\savesymbol{tablenum}
\usepackage{siunitx}
\restoresymbol{SIX}{tablenum}
\defcitealias{2020MNRAS.492.6105M}{M20}
\DeclareMathOperator\erf{erf}
\hypersetup{linkcolor=blue,citecolor=cyan,filecolor=green,urlcolor=magenta}

\received{2021-2-17}
\revised{2021-4-9}
\accepted{2021-4-14}
\submitjournal{AJ}

\shorttitle{Track Before Detect Earth Trojan Survey}
\shortauthors{Lifset et al.}

\begin{document}

\title{A Search for L4 Earth Trojan Asteroids Using a Novel Track-Before-Detect Multi-Epoch Pipeline}

\correspondingauthor{Noah Lifset}
\email{nlifset1@gmail.com}

\author[0000-0003-3397-7021]{Noah Lifset}
\affiliation{Lawrence Livermore National Laboratory, 7000 East Avenue, Livermore, CA 94550, USA}

\author[0000-0003-2632-572X]{Nathan Golovich}
\affiliation{Lawrence Livermore National Laboratory, 7000 East Avenue, Livermore, CA 94550, USA}

\author[0000-0003-4338-8095]{Eric Green}
\affiliation{Lawrence Livermore National Laboratory, 7000 East Avenue, Livermore, CA 94550, USA}

\author[0000-0002-6911-1038]{Robert Armstrong}
\affiliation{Lawrence Livermore National Laboratory, 7000 East Avenue, Livermore, CA 94550, USA}

\author[0000-0002-2582-0190]{Travis Yeager}
\affiliation{Lawrence Livermore National Laboratory, 7000 East Avenue, Livermore, CA 94550, USA}

\begin{abstract}

Earth Trojan Asteroids are an important but elusive population that co-orbit with Earth at the L4 and L5 Lagrange points. There is only one known, but a large population is theoretically stable and could provide insight into our solar system's past and present as well as planetary defense. In this paper, we present the results of an Earth Trojan survey that uses a novel shift-and-stack detection method on two nights of data from the Dark Energy Camera. We find no new Earth Trojan Asteroids. We calculate an upper limit on the population that is consistent with previous searches despite much less sky coverage. Additionally, we elaborate on previous upper limit calculations using current asteroid population statistics and an extensive asteroid simulation to provide the most up to date population constraints. We find an L4 Earth Trojan population of $N_{ET} < 1$ for $H = 13.93$, $N_{ET} < 10$ for $H = 16$, and $N_{ET} < 938$ for $H = 22$.
\end{abstract}


\section{Introduction}

\subsection{Earth Trojan Asteroids}

Earth Trojan Asteroids (ETAs) are an important population of inner Solar System asteroids that could unlock a better understanding into the development and current epoch of the Solar System. They co-orbit the Sun with the Earth at the forth and fifth Lagrange points (L4 and L5) as solutions to the reduced three-body problem \citep[see e.g.][]{1967torp.book.....S}. ETAs have been estimated to be stable on the order of 100 Myr to 1 Gyr with some specific orbital regimes offering stability of up to the age of the Earth \citep{2012MNRAS.426.3051C, 2013CeMDA.117...91M, 2019A&A...622A..97Z}. The oldest and most stable ETAs, referred to as primordial ETAs, could be remnants from the protoplanetary disk, and characterizing them could inform our understanding of the primordial Solar System and its development. Studying ETAs can also provide insight on the Earth and Moon. A leading lunar formation theory is the giant impact hypothesis between proto-Earth and Theia, which may have been a Mars-sized ETA that outgrew the stability of the L4 Lagrange point \citep{2005AJ....129.1724B}. If proven, this would lend substantial credence to primordial ETAs. Along with near-Earth steroids, ETAs could explain the asymmetry in lunar impact craters between the leading and trailing hemispheres \citep{2010AGUFM.P53C1537I}. ETAs are also prime targets for space missions, because they have a low relative velocity \citep{2019NatAs...3..193M}. This also means that if they are destabilized by resonances or other perturbations, they could endanger life on Earth. 

While Trojan asteroids have been discovered for Venus \citep{2014MNRAS.439.2970D}, Mars \citep{2013MNRAS.432L..31D}, Jupiter \citep{2017AJ....154...71Y}, Uranus \citep{2015MNRAS.453.1288D} and Neptune \citep{2009A&A...508.1021A} only one ETA has been found to date. This is likely due to the difficulty in detecting them as they spend much of their trajectories near the Sun on the sky. ETAs co-orbit with Earth 60 degrees ahead or behind (on average), corresponding to L4 and L5 respectively. As a result of this geometry, ground based observations are limited to mostly twilight observations and only short dark-sky observations depending on the season \citep{1998Icar..136..154W}. The geometry also ensures a poor reflectance angle with the Sun, so any objects in these orbits are significantly fainter than analogous asteroids at opposition.

The only ETA found to date, 2010 TK7, was a serendipitous detection by WISE \citep{2000DPS....32.1407C} at a portion of the orbit that brought it close to Earth with a large solar elongation. Subsequent follow up suggests it was recently captured and is relatively unstable. Its lifetime is estimated to be $\sim$7-250 thousand years \citep{2011Natur.475..481C, 2012A&A...541A.127D}. This type of quasi-stable, captured orbit is distinct from theoretically long-term stable primordial ETAs, which lie in slightly different orbital regimes. These stable ETAs have lifetimes estimated to be as large as Earth's age, although the Yarkovsky effect could potentially limit this for smaller asteroids \citep{2019A&A...622A..97Z}. The asteroid 2010 TK7 is somewhat of an outlier and separate from the target population of this paper: long term stable, primordial ETAs.

\subsection{Previous Searches}

There have been four previous searches for ETA populations. All of them found none, but three have offered a population constraint. \cite{1998Icar..136..154W} used a ground based survey to search L5. They covered 0.35 square degrees and set a constraint of $3$ objects per square degree at $r = 22.8$, corresponding to 350m (175m) for C (S) class asteroids. \cite{2018LPI....49.1149C} searched L4 using the OSIRIS-REX spacecraft on a flyby on its way to the asteroid Bennu. They covered nine 16 deg$^2$ patches and calculated an upper limit of $73 \pm 23$ at $H = 20.5$, which corresponds to 470/210m for C/S class asteroids. They also applied their analysis to the data from \citet{1998Icar..136..154W}, resulting in an upper limit of $604 \pm 358$ at $H = 21.2$ $(r = 22.8)$ and $194 \pm 116$ at $H = 20.5$. The Hayabusa2 spacecraft searched L5 on its way to the Ryugu asteroid \citep{2018LPI....49.1771Y}, but published no ETA population constraint analysis beyond a null detection. \cite{2020MNRAS.492.6105M} searched for ETAs at L5 using the Dark Energy Camera. They covered 24 square degrees and calculated an upper limit of $N_{ET} < 1$ at $H = 15.5$, $N_{ET} < 60 -- 85$ at $H = 19.7$, and $N_{ET} < 97$ at $H = 20.4$; this is the strictest constraint on the ETA population to date. \linebreak

In this paper we present the results of a new search for ETAs at the Earth-Sun L4 point and the new population upper-limit calculated from a null result. In \S\ref{sec:data}, we describe our observations and reduction. In \S\ref{sec:methods}, we describe the detection pipeline and the ETA simulation we used to determine our upper limits. In \S\ref{sec:results}, we present the results of our detection pipeline and the upper limit calculation. In \S\ref{sec:discussion}, we compare our results to previous searches and provide a meta-analysis of ETA population upper limits across the literature.

\section{Data} \label{sec:data}

\subsection{DECam Observations}

Our survey consisted of roughly 90 minutes of data on both 2019 Nov 20 and 21 on the Blanco telescope, which is located at Cerro Tololo Inter-American Observatory in Chile, using the Dark Energy Camera \citep[DECam][]{2008arXiv0810.3600H}. DECam has a 4 meter diameter mirror and 62 science active 4k$\times$2k CCDs (only 60.5 currently useful) with 0$\arcsec.263, \text{pixel}^{-1}$. We observed in the r-band, choosing slightly lower solar reflectance from potential ETAs for lower sky background during twilight. We unfortunately failed to include dithers in the observing scripts. However, since our detection pipeline stacks signal across all exposures, we are still sensitive to ETAs that traverse chip-gaps during our survey time. 

Night 1 and 2 consist of 78 and 77 exposures, respectively, with all exposures from a given night stacked on the same position. The exposures were all 40 seconds, with roughly 30 seconds between exposures for readout. The exposure time and number of images per night were chosen as a balance between a number of factors, such as more images improve the median stack for difference imaging, fewer images decrease the number of times read noise is added, more images allow for averaging over noise frame to frame, shorter exposures can decrease trailing losses, and the limited amount of potential observation time for ETAs. We covered an area of 5.72 deg$^{2}$ located near the L4 Lagrange point (see Figure \ref{fig:observations}). This effective area is estimated by a computing the area of a circle of radius 1.1$^\circ$, while accounting for the overlap in pointing between the two nights. We selected a location slightly east of the Lagrange point because the peak of the on sky spatial distribution is located there as a result of the angled observation geometry and the asymmetry of tadpole Trojan orbits. ETAs are expected to have a motion of $\sim1^\circ\,\text{day}^{-1}$. This allows for consecutive-night detections for typical ETA trajectories on sky. 

\subsection{Reduction Pipeline}

\begin{figure}
  \centering
  \includegraphics[width=\columnwidth]{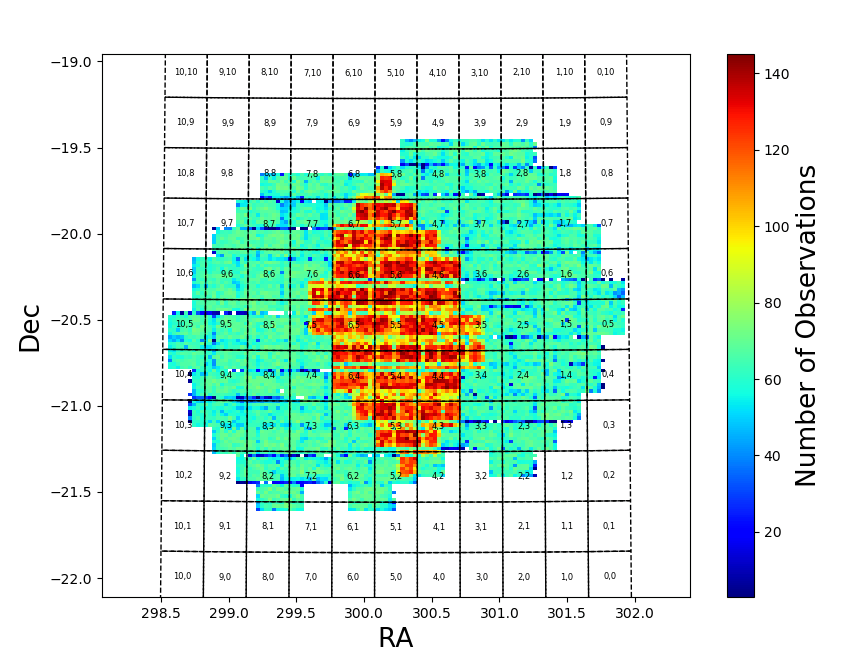}
  \caption{The data set depth (number of frames) in R.A. and Dec. We cut the entire region into ``patches'' with the LSST data management pipeline, followed by creation of a master pixel grid. All analyses occurred on this grid.}
  \label{fig:observations}
\end{figure}

We obtained our exposures after standard processing by the DECam Community Pipeline \citep{2014ASPC..485..379V}. We then used the LSST software package\footnote{https://pipelines.lsst.io} \citep{2019ASPC..523..521B} to produce difference images. This can be split into three different pieces: making calibrated single epoch catalogs and images, creating a static template image, and subtracting the template from the single epoch images to produce difference images.

The LSST stack single epoch processing includes: the masking of cosmic rays, measuring the PSF, detecting objects, deblending and measuring individual sources and calibrating the astrometry and photometry. A detailed explanation of each of these steps is explained in the HSC data release papers \citep{2018PASJ...70S...8A,2019PASJ...71..114A}. For photometric calibration, we used the Pan-STARRS catalog \citep{2020ApJS..251....7F}.

To create the template images we first interpolated the data onto a tangent plane projection centered on the average pointing. The total observed area was divided into a regular grid with each grid $4000\times4000$ pixels or $\sim 17\arcmin$ on a side (see Figure \ref{fig:observations}). For the templates we selected all images with PSF full-width at half maximum $\text{seeing} < 1.2\arcsec$. For each grid the static sky is constructed following the procedure outlined in \citep{2018PASJ...70S...8A,2019PASJ...71..114A}. Briefly, the procedure builds a two-sigma clipped coadd to construct a static image of the sky. We subtract this coadd from each individual image and identify variable sources. Those detections that are truly variable will only appear in a small subset of the single epoch visits. The variable sources are then masked and a coadd image is created by taking the mean of all the images.

We use the Alard-Lupton algorithm \citep{1998ApJ...503..325A} as implemented in the LSST stack to create difference images. This procedure estimates a convolution kernel which, when convolved with the template, matches the PSF of the template with that of the science image by minimizing the mean squared difference between the template and science image. The Alard-Lupton procedure uses linear basis functions, with potentially spatially-varying linear coefficients, to model the matching kernel which can flexibly account for spatially-varying differences in PSFs between the two images. The algorithm has the advantage that it does not require direct measurement of the images’ PSFs. Instead it only needs to model the differential matching kernel in order to obtain an optimal subtraction. 

After examining the difference images, we found that there remained a large number of artifacts due primarily to three issues: WCS, PSF matching and bright stars. The largest issue we found was the accuracy of the WCS from exposure to exposure causing stars to be dipoles. We were unable to improve the WCS by combing multiple exposures due to the small dithers. We used the best seeing images to create the static sky template, but didn't exclude these images from our analysis. Therefore, there were some images that required a de-convolution when matching the PSF. Finally, bright stars and galaxies can cause problems in the difference images because our PSF model is not sufficiently accurate in the wings and therefore bright objects are often not subtracted well. We tried to remove some of these issues by masking out known objects. We selected objects from PAN-STARS with $m_r < 20$ and masked them using the procedure described in \citet{2018PASJ...70S...7C}. For each star the mask was composed of a magnitude-dependent circle for the star and a rectangle for the bleed trail. We visually tuned the size of the circle and rectangle to match our data. We split the data into two class above and below 14th magnitude. The size of the radii in arcseconds are: 
\begin{equation}
  {\rm radius} = 
    \begin{cases}
     400 \, e^{-m_{r}/3.8}, & \text{for}\ 14<m_{r}<20 \\
     400 \, e^{-m_{r}/4.1} , & \text{for}\ m_{r}<14. 
    \end{cases}
\end{equation}The length and width of the rectangles are 1.5$\times$0.15 times the circle's diameter and for $14<m_r<20$ and 6$\times$0.3 times the diameter for $m_r < 14$. The longer rectangles were necessary to remove some long bleed trails for bright stars.

\section{Methods} \label{sec:methods}

We employ a shift-and-stack method to detect asteroids in the data. By combining the signal from multiple images, we can reduce the noise floor and increase sensitivity. We use a novel shift-and-stack method developed by \cite{2021arXiv210403411G} that can be described as track-before-detect. This works by generating a large number of random potential asteroid trajectories within the search criteria phase space, calculating the signal-to-noise-ratio (SNR) for each trajectory in the data, and saving the information of potential detections if the SNR meets a given threshold. This process is then repeated until the phase space of the search criteria has been completely filled through the random trajectory generation. 

Traditional shift-and-stack methods will usually stack entire images on top of each other according to a selected angle or proper motion \citep[e.g.][]{2015AJ....150..125H}. This can work just as well and will save some computational time since all parallel trajectories are simultaneously sampled by detecting on the shifted stack. However, this is restricted to linear source motion, because different proper motions can not be applied to different parts of the same stacked images. Track-before-detect allows for non-linear trajectories at the (substantial) cost of sampling trajectories individually, because all that is needed is the positions of intersections between a given trajectory and images in the data set. This method has been proposed before as ``nonlinear shift-and-stack" by \citet{2010PASP..122..549P}. In this paper, we use linear trajectories, which were sufficient to generate competitive results in an ETA constraint; however, we implement our track-before-detect pipeline as a second step (following \cite{2021arXiv210403411G}) toward non-linear trajectories and applications to larger data sets spanning much more time and sky coverage. In general, track-before-detect is most useful in cases involving fast moving asteroids whose nonlinear motion manifests within one one night, data sets that span large amounts of time where normal asteroids show nonlinear motion, or both. 

\subsection{Detection Pipeline}
\begin{figure}
  \centering
  \includegraphics[width=\columnwidth]{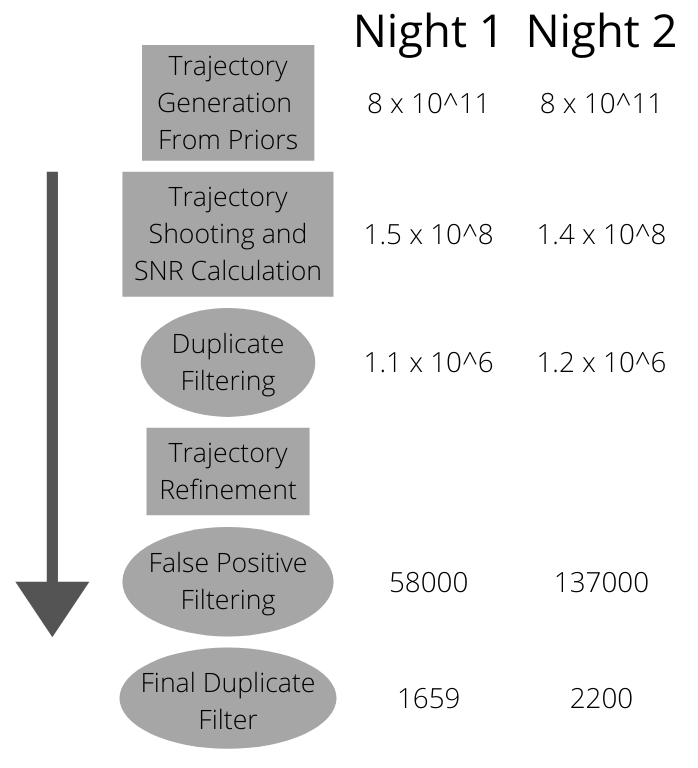}
  \caption{A flow chart demonstrating the detection pipeline from start to finish. Ovals represent filters and boxes represent calculations or trajectory interactions. Numbers of potential detections after each step are shown on the right for both nights. The final potential detection lists are manually sorted to produce final detection lists. Differences between the two nights are due to variation of difference imaging artifacts, sky noise variation, and inherent randomness.}
  \label{fig:flow_chart}
\end{figure}

The detection pipeline is shown as a flow chart in Figure \ref{fig:flow_chart} with the number of potential detections shown at each step for both nights. We run our detection algorithm separately on each night with the same search criteria phase space, because ETAs can not be expected to move linearly on time scales that would include both nights of data. We represent each night's survey data in memory as a virtual, linear, three-dimensional coordinate space that encompasses the minimum bounding box to contain the survey's image data. In this space, the X- and Y- axes correspond to individual pixels in the image data while the Z-axis represents time in seconds. This also allows us to virtually stitch together the images from each exposure by placing the individual image data at the correct positions within this three-dimensional data space. Each exposure represents 40 seconds of real time, so pixels may be imagined as rectangular prisms with cross-sectional area equal to the pixel area, and height proportional to the exposure time

The pipeline begins with trajectory generation within the defined search criteria phase space. Each trajectory is defined by 5 numbers: X-intercept, Y-intercept, Z-intercept/time, X-slope, and Y-slope. The X- and Y-intercepts are the pixel position of the trajectory at that Z-intercept/time and the X- and Y-slopes are the corresponding velocities in pixels s$^{-1}$. In trajectory generation, the initial values of the X-, Y- and Z-intercepts were chosen by randomly selecting a non-NaN pixel in the entire night's data and propagating back to the beginning of the first exposure to calculate intercept values. The slopes were randomly generated from a uniform rectangle in ecliptic coordinates: -40 to 40 $\arcsec$ hour$^{-1}$ in longitude and 110 to 190 $\arcsec$ hour$^{-1}$ in latitude. This range was chosen based on \citet{2000Icar..145...33W} to cover nearly all potentially observable ETAs and aligns with the range used by \citet{2020MNRAS.492.6105M} (0.75--1.25 $^\circ\,\text{day}^{-1}$). We generated $8 \times 10^{11}$ trajectories for each night in order to fill the parameterized space. This number was chosen empirically as the recovery rate of injected asteroids no longer improved with additional trials. 

\subsubsection{Pixel Values and Memory Management}

Once the trajectories are generated, we calculate their intersections with each of the images in the data set and pull the pixel values for each of those points. However, the large number of trajectories and the full volume of image data make this process complicated in practice. We first split the total number of trajectories into equally-sized work units to be distributed among worker nodes on our computing cluster. These work units are further divided into a series of batches by the workers. For each night of data, we tested approximately 824 billion trajectories across 768 work units. Each work unit evaluated approximately 536 million trajectories that was further subdivided into 8 batches of approximately 67 million individual trajectories drawn from the motion model search criteria.

Each batch is processed by sequentially streaming through all of the exposures in the survey data and collecting the information needed from each image. For every intersection point in an exposure, we calculate a weighted sum of the image pixels and store the local flux, the number of non-NaN pixels intersected, and local SNR (see \S\ref{subsubsec:SNR} below). 

Streaming through the data in this manner is bandwidth-intensive and results in a balancing act between the number of exposures to store in memory and the number of trajectories in each batch. Larger batch sizes directly increase the disk efficiency of the process because more floating point operations are done at each step, but this leaves less available space to pre-load data from disk, potentially leading to stalling while waiting for data from disk. The maximum size of each batch is constrained by the total amount of memory available to the worker and the number of pre-loaded exposures.  

Our approach is to attempt to read in the next five exposures from disk while the active exposure is being processed and leave the rest of the available memory free for storing the intermediate results for each batch. In most cases this allows us to avoid waiting for disk accesses because we are able to read in all of the data for the next exposure before the current batch has finished processing. In this manner we are able to access the sizable processing power of our compute clusters while also avoiding the data access delays. Furthermore, we are able to replicate the jobs in an embarrassingly parallel fashion on the requisite number of compute nodes in order to achieve our desired completeness. Our detection pipeline was run on the Catalyst cluster at Lawrence Livermore National Laboratory\footnote{https://computing.llnl.gov/computers/catalyst}.

\subsubsection{SNR Calculation}\label{subsubsec:SNR}

Once the requisite pixels are pulled from the data, the SNR is calculated with a weighted sum on the optimal signal matched filter using a two-dimensional Gaussian profile convolved with a line segment equal to the length of the expected streak, which we can compute from the sampled proper motion for each trial trajectory and the exposure time. The flux and noise in each pixel were summed across all images that a given source intersected. Within one image intersection, we measure the flux $F$ by calculating a weighted sum over the pixels $(p)$ inside the streak: 
\begin{equation}
  F = \sum_{i}{w_i\,p_i}. 
\end{equation}The weights $(w)$ are determined from the profile, which is the convolution of a line segment with a two-dimensional Gaussian. The weights are normalized such that their sum equals the effective number of pixels in the streak: 
\begin{equation}
  n_{eff} = \sum_{i} w_i. 
\end{equation}The effective number of pixels is calculated by integrating the flux profile:
\begin{equation}
  n_{eff}^{-1} = \int{dxdy\,P(x,y)^2}. 
\end{equation}Specifically, we consider a streak with length $L$ oriented along the x-axis, width $\sigma_y$, flux per unit length $l_0$, and center (0,0), convolved with a symmetric bivariate Gaussian PSF of width $\sigma_{\pi}$: 
\begin{multline}
  P(x,y) = \frac{l_0}{2L} 
  	\frac{1}{ \sqrt{2\pi(\sigma_{\pi}^2 +\sigma_y^2)} } 
  	\exp\left(
  		\frac{-y^2/2}{\sigma_{\pi}^2+\sigma_y^2}
  	\right) \,
  	\times \\
  	\left(
  		\erf{\left(
  			\frac{L - 2x}{2\sqrt{2\sigma_{\pi}^2}}
  		\right)} + 
  		\erf{\left(
  			\frac{L + 2x}{2\sqrt{2\sigma_{\pi}^2}}
  		\right)}
  	\right).
\end{multline}
Assuming that the object is unresolved, $\sigma_{\pi} >> \sigma_{y}$, this simplifies accordingly:
\begin{multline}
  P(x,y) = \frac{l_0}{2L}\frac{1}{\sqrt{2\pi\sigma^2}}\exp{\left({\frac{-y^2}{2\sigma^2}}\right)}
\,\times \\\left(\erf{\left(\frac{L - 2x}{2\sqrt{2\sigma^2}}\right)} + \erf{\left(\frac{L + 2x}{2\sqrt{2\sigma^2}}\right)}\right),
\end{multline}where we have defined $\sigma \equiv \sigma_{\pi}$.

We choose an area surrounding the streak to sum over that balances capturing as much signal as possible with limiting computation time. We integrate flux to $3\sigma$ assuming a Gaussian point spread function away from the streak's ridge, which encompasses over 99$\%$ of the signal.

Sources of noise include sky background Poisson noise, read noise, dark current, and Poisson noise from the signal itself. The total noise $N$ includes each of these but is dominated by the sources of Poisson noise. The noise sources are sufficiently large enough to be added in quadrature as Gaussian random variables. Thus we have:
\begin{equation}
  SNR = \frac{F}{\sqrt{F + N}}.
\end{equation}The background noise is calculated as the standard deviation of pixel values in the difference image. The expected value per pixel will then be the square of this number, and thus we get the total background signal as
\begin{equation}
  N = \sigma_{\rm pixel}^2\, n_{\rm eff}. 
\end{equation}From Equations 2-6, the flux from the asteroid is given by
\begin{equation}
  F = \frac{\sum_{i}{(l_i \, p_i)} \, \sum_{i}{l_i}}{\sum_{i}{l_i^2}},
\end{equation}where the sum over pixels approximates the integral.

There is a chance that a valid detection might overlap on one image with an image artifact like a cosmic ray. To compensate for this, we store an initial light curve for each trajectory consisting of the local SNR value on each intersected image. We then apply an outlier rejection method, consisting of a rolling median sigma clip, to this temporary light curve and ignore those clipped images when calculating the final SNR for the trajectory. We store information on any trajectory that has an SNR greater than 5 and intersects at least 5 images. Most of these (see Figure \ref{fig:flow_chart}) are false positives or duplicate detections of the same object that will later be filtered out.

\subsubsection{Detection Post-processing}

Once we have initial detection lists for each night, we apply a filter to remove duplicate detections of the same objects. This is common for bright asteroids, because a trajectory that is only partially aligned with the asteroid will still surpass the SNR threshold. We use a k-d tree to find nearest neighbors for all the trajectories in the four dimensional space of $(x,y)$ position and proper motion, where we convert the proper motions to spatial coordinates by multiplying by the total integration time of the data for each night. Using this four-distance metric, we consider all trajectories within $\epsilon=5\arcsec$ of others duplicates. To be explicit, we require:
\begin{equation}\label{eq:4dist}
  \sqrt{\left(|\vec{x}_{0,i}-\vec{x}_{0,j}|\right)^2 + \left(\Delta t \, |\vec{\pi}_i-\vec{\pi}_j|\right)^2} < \epsilon, 
\end{equation}where $\vec{x}_{0,ij}$ are the position of two trajectories at the first exposure, $\Delta t$ is the time between the last and first exposure, and $\vec{\pi}_{ij}$ are the proper motions. We keep the trajectory with the highest SNR. We chose $\epsilon=5\arcsec$ as a balance between removing a meaningful number of duplicates and being small enough to have almost no chance of removing any two different objects that are close by coincidence. This cut is small enough that many duplicate trajectories will still pass through the filter, which will be addressed by a stronger duplicate filter at the end of the pipeline.

We then refine each trajectory with a \textsf{Scipy} \citep{2020SciPy-NMeth} optimizer using the Nelder-Mead method to maximize global SNR. We save small image cutouts of each trajectory-image intersection and use the same SNR calculation as above on these image cutouts as an optimization function. We then take these refined trajectories and apply a few stronger filters to remove potential false positives. 

First, we apply a series of filters on the median stack image. This median image is calculated by taking the median across the image cutouts for each trajectory. In this median image, a trajectory that perfectly aligns with a source would appear as a centered single PSF (or short streak for fast moving sources). We calculate the SNR of this median stack using a the same weighted sum as before and we require $SNR_{med}>7$. Next we calculate the the $\chi^2$ of the center of the median image and the $\chi^2$ of an annulus region around it. To be explicit, we calculate:
\begin{equation}
  \chi^2 = \frac{1}{N_{pix}}\sum{\frac{p_i^2}{N_i^2}}
\end{equation} where the center consists of pixels within $3\sigma$ of the center and the annulus consists of pixels between $5\sigma - 10\sigma$, where $\sigma$ is with width of the Gaussian PSF. We require that $\chi_{annulus}^2 < 2$ or $\chi_{center}^2 < \chi_{annulus}^2 - 1$. The idea behind this filter is to discard false positives based on shape. False positives that result from difference image artifacts usually have signal in the annulus region as well as the center. The exact cut was determined empirically by calculating the $\chi^2$ values for a set of known injected asteroids and setting limits that would balance discarding the false positives with not discarding any valid detections.

\begin{figure}
  \includegraphics[width=\columnwidth]{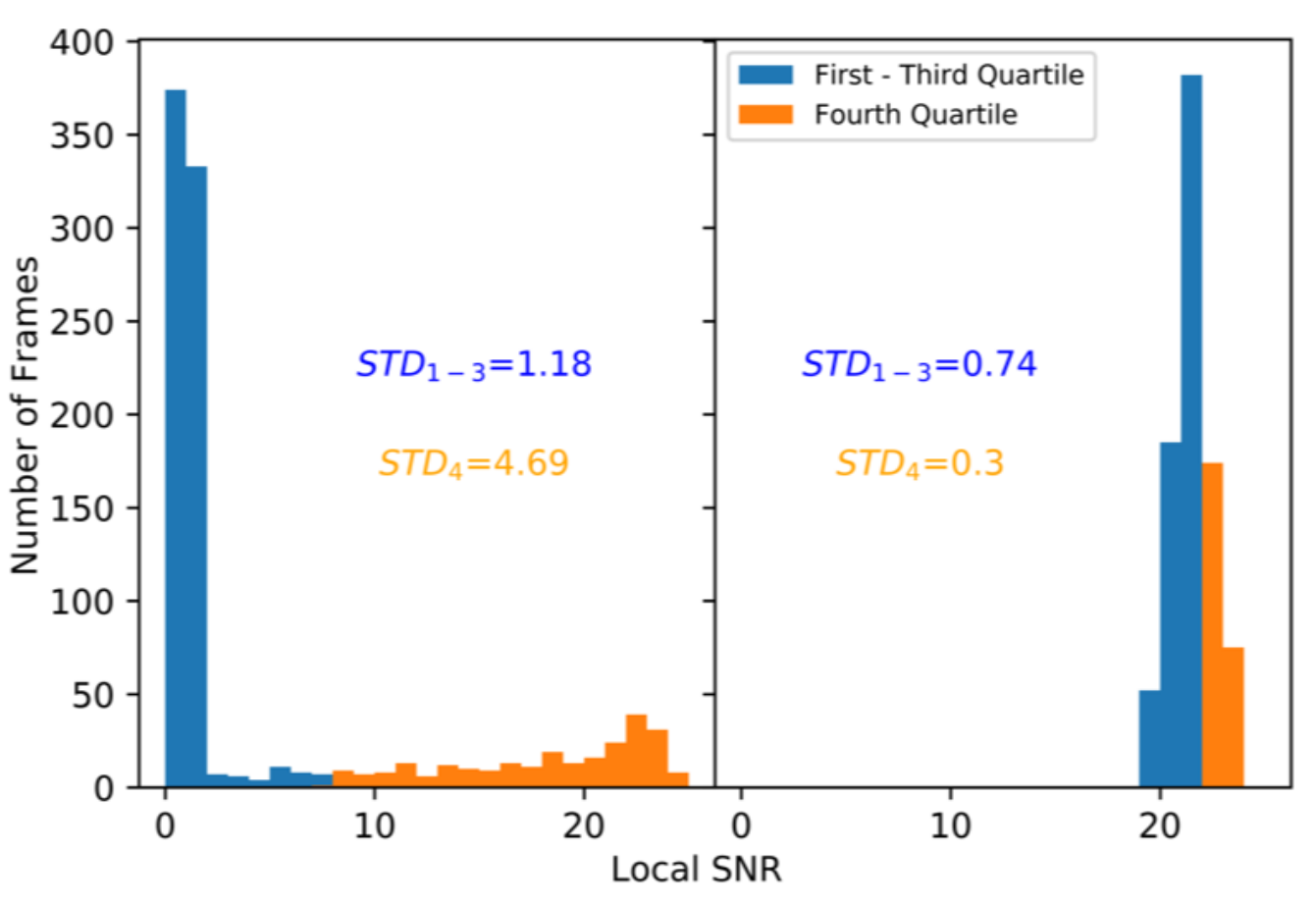}
  \caption{Example distributions of Local SNR values of frames hit for detection trajectories that are bad (left panel) and good (right panel) approximations of the true trajectory. The near hit filter compares the standard deviation of the first through third quartile (blue bins) with that of the fourth quartile (orange bins). The near-hit mostly has local SNRs around 0 and some reaching higher corresponding to when the trail trajectory overlaps the flux from the asteroid, and so fails the test. The right panel shows a good hit, which has mostly non-zero local SNRs distributed around a typical value related to the brightness of the asteroid.}
  \label{fig:nearhit_filter_example}
\end{figure}

Next, we apply a ``near-hit'' filter that attempts to cut out any trajectory that only partially overlaps with a real object in the data. These are often duplicate detections of bright objects as previously discussed, but are too far in 4-distance from the center to be filtered out that way. The near-hit filter is carried out by analyzing the distribution of local SNR values in the light curve. For a good hit, most of the local SNR values would distribute around a mean value due to Poisson noise in the flux and the sky background as well as intrinsic variability due to rotation of the asteroid. For a near-hit, the portion of the trajectory that misses the true asteroid motion would exhibit near-zero SNR. To differentiate between these two, we rank-order the local SNR values into an array and compared the standard deviation of the highest values of the distribution (fourth quartile) to the standard deviation of the lowest (first through third quartile). We required that the fourth quartile variance be less than the first through third quartile variance, thus requiring that most of the high value local SNRs are clumped around a value indicative of a detection on all of the frames that intersect the trajectory. We demonstrate asteroids that both pass and fail this cut in Figure \ref{fig:nearhit_filter_example}. In the case where all values are clumped around zero, and this filter is passed, the overall SNR would likely be below our threshold in the first place. We also apply a similar filter where we take each trajectory and randomly split its light curve in two. We then require that each half individually must reach the SNR cut of $\frac{5}{\sqrt{2}}$ to check that flux is roughly split between two random halves of the light curve. 

After false positive filtering, we apply another stronger duplicate filter with a 4-distance cutoff of $30 \arcsec$. This was chosen because it is the same 4-distance used for matching detections with known objects like MPC catalog asteroids or injected false asteroids (see \S\ref{subsec:detections}). This final duplicate filter reduces the potential detection list to a small number that can be manually sorted and ultimately made into a final detection list.

\subsection{False Source Injection}

In order to test completeness, we create an identical data set with false asteroids injected during data reduction. Roughly 4000 false asteroids were injected separately in each night. They were injected as linear trajectories with slopes randomly selected from the same search criteria phase space used in detection. Their X-, Y-, and Z-intercepts were generated by randomly selecting a valid right ascension, declination, and time within the bounds of the night's data. They had apparent magnitudes randomly selected between 17 and 26. Any potential injections that intersect fewer than five images were not considered in the completeness calculation. 

\begin{figure}
  \includegraphics[width=\columnwidth]{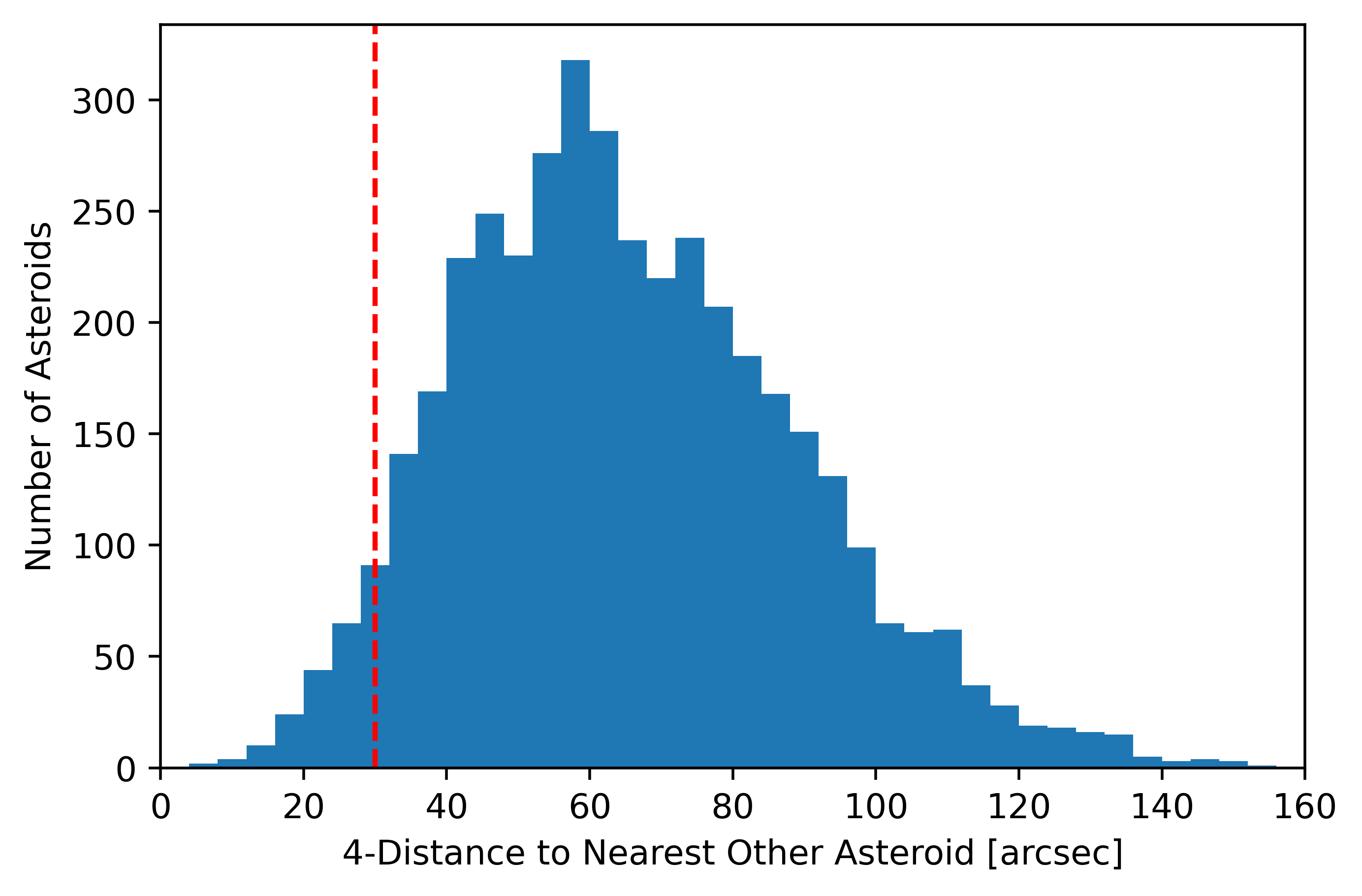}
  \caption{Distribution of 4-distances from each injected asteroid to the closest other injected asteroid. We show this for each asteroid instead of each closest pair, so all numbers are double counted. A cutoff of $30 \arcsec$ was used despite the potential to rule out two genuinely separate injected asteroids, because it was needed to include nearly aligned detections in order to reduce overall potential detections more significantly.}
  \label{fig:self_diff}
\end{figure}

We determined which detections corresponded to injected asteroids using the 4-distance metric defined in Equation \ref{eq:4dist}. A given injected asteroid was considered matched with a detection if there was one within a 4-distance of $30 \arcsec$ and the closest one was selected if there were multiple. We chose this cutoff empirically. We found that $30 \arcsec$ was large enough to include nearly all valid detections including those that did not align perfectly. It was small enough to rarely include two actually separate injected asteroids that were by chance overlapping (see Figure \ref{fig:self_diff}). Once matches are made we are able to measure completeness as a function of apparent magnitude.

\subsection{ETA Simulation}

In order to calculate a constraint on the ETA population, we determine the fraction of a theoretical ETA population that our observations cover. This is done by simulating the orbits of millions of asteroids over \SI{2.2e7} years to determine the spatial distribution of long-term stable ETAs. The simulation only focuses on long-term stable ETAs and not captured ETAs, such as 2010 TK7. So the fraction of the theoretical ETA population calculated and resulting constraint only apply to long-term stable ETAs and not captured ones. One of the main improvements this paper makes over previous ETA population calculations is the use a large scale simulation in determining the distribution of stable ETAs for the constraint calculation. 

\begin{figure}
  \includegraphics[width=\columnwidth]{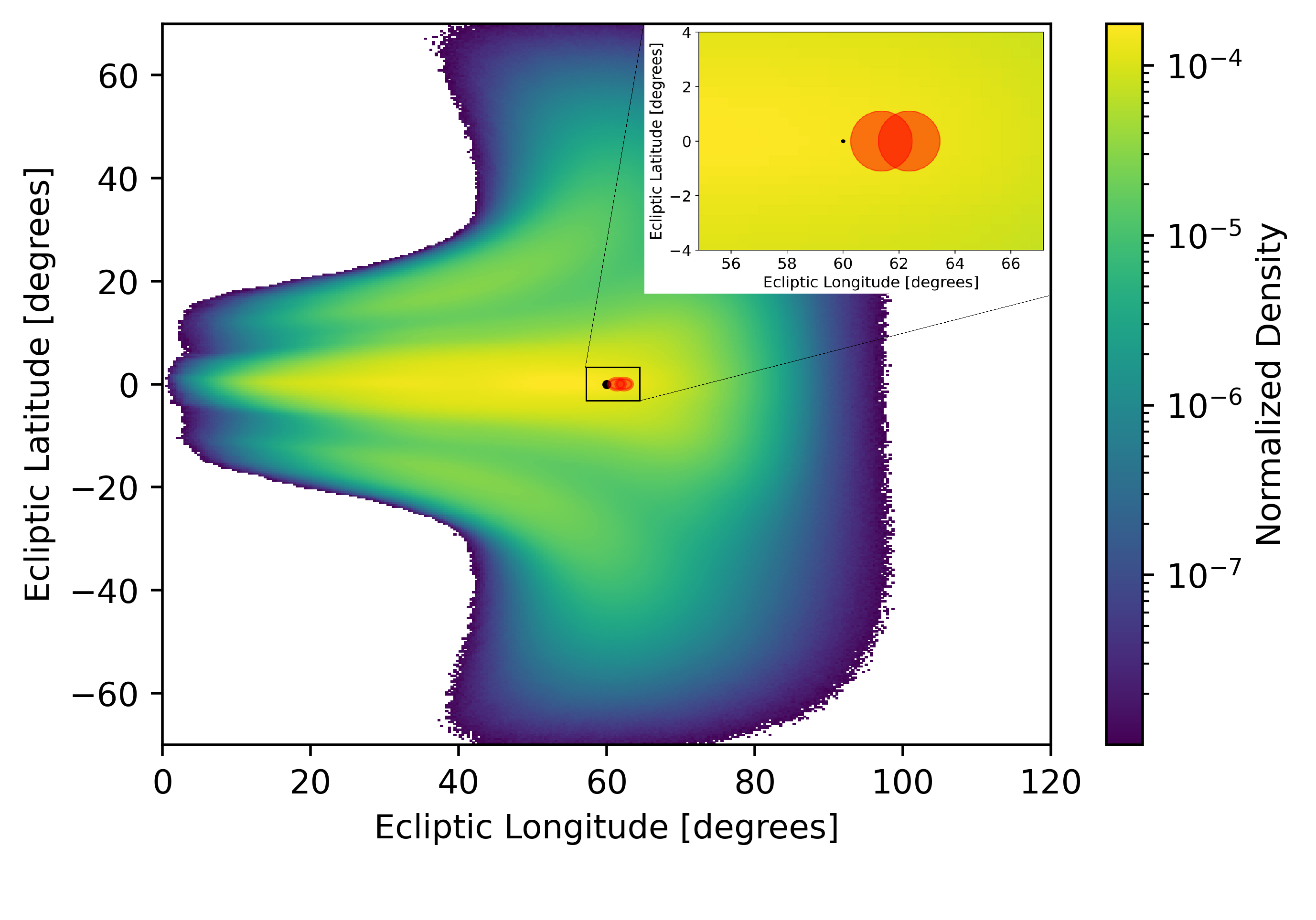}
  \caption{Spatial distribution of stable ETA population. Points are sampled every 10 years for 1 million years and then every 1,000 years after to a total of 22 million years. The black dot represents the Lagrange point and red circles the two nights of observations. Positions at each time step are included in order to show the time average position of ETAs.}
  \label{fig:ETA_sim}
\end{figure}

The simulation was run with the \textsf{Rebound} N-body integrator software using a IAS15 integrator with adaptive time steps \citep{rebound, reboundias15}. We initialized roughly 8 million point particles in a simulation including the Sun, the Earth, and all other planets. Inclinations were randomly selected from a Gaussian distribution centered on 0 with a standard deviation of 10 degrees. Eccentricities were randomly selected from a Gaussian distribution centered on 0 with a standard deviation of 0.0667 (absolute value taken after sampling). Semi-major axes were randomly selected from a Gaussian distribution centered on 1 AU with a standard deviation of 0.0667 AU. All other orbital elements were randomly selected from a uniform distribution from 0 to $2\pi$. If at any point in the simulation, a point particle moved outside the half-sphere on the L4 side of the orbit relative to the plane perpendicular to the ecliptic that contains the Earth--L3 line, it was no longer considered a stable ETA and was discarded. This criteria was focused on finding long term stable ETAs including large tadpole orbits but not horseshoe orbits that continue beyond L3 to the other half-sphere. Of the initial 4 million point particles with Earth-like orbits, the simulation ended with 38,000 stable ETAs under the influence of the Earth and the Sun. Positions are sampled every 10 years for the first 1 million years and then every 1,000 years after that. Each sample of stable ETAs that last until the end is included to show time average positions. The geocentric ecliptic coordinates are calculated and shown in Figure \ref{fig:ETA_sim}. 

\section{Results} \label{sec:results}

\subsection{Detections}\label{subsec:detections}

\begin{figure}
  \includegraphics[width=\columnwidth]{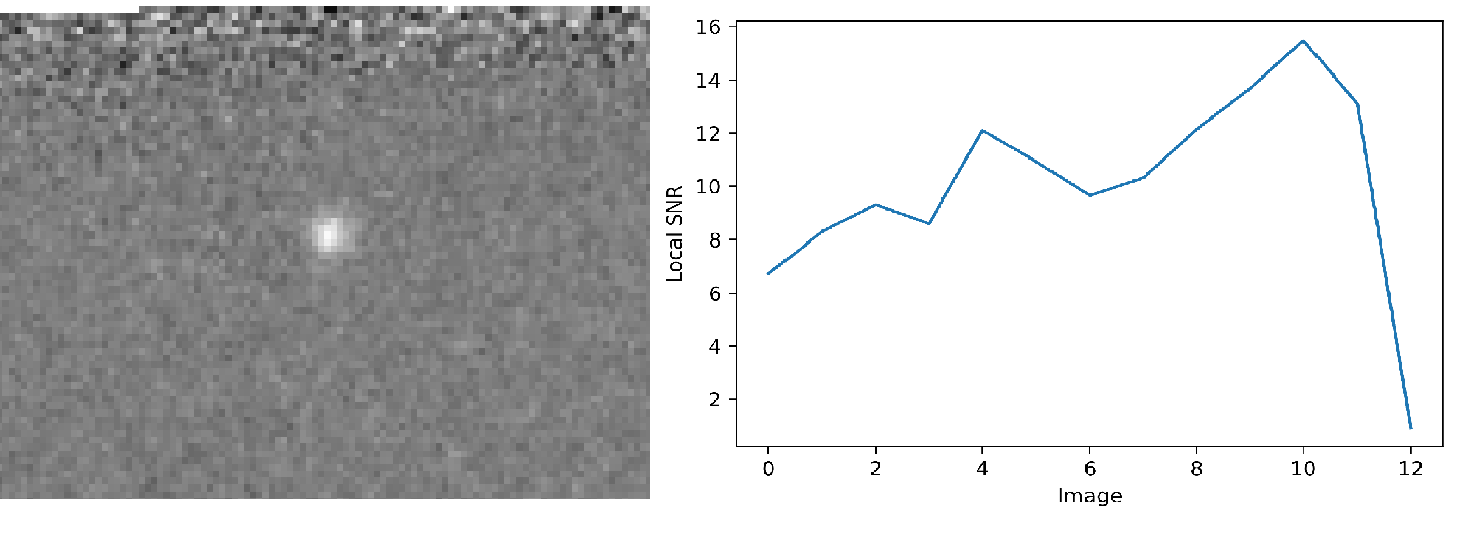}
  \caption{Left: Median stack image of the new detected asteroid. The increased noise at the top is because the CCD edge is nearly aligned with the top of this cutout. Right: light curve of new detected asteroid with local SNR as a function of image intersected. The drop at the end is from a star mask that overlaps with the trajectory.}
  \label{fig:new_ast}
\end{figure}

We found no ETA candidates in either night of observations. Our search criteria phase space was strict and focused on potential ETAs, so we expected to detect few objects in general. We detected 55 objects: 18 on night 1 and 37 on night 2. All except one were known objects in the MPC catalog. The single new detection was found to have a proper motion of 0.58$^\circ\,\text{day}^{-1}$, which was outside the expected range for ETAs. It is likely that we detected this object, as well as the others, on the edge of our search criteria phase space and the refinement process aligned it more properly outside the phase space. Its median image and light curve are shown in Figure \ref{fig:new_ast}. We detected four Mars-orbit crossing Amor type asteroids and one Earth-orbit crossing Apollo type asteroid. Our ability to detect an Apollo asteroid and four Amor asteroids is a testament to our pipeline's ability to detect asteroids of interest in the same areas that an ETA would exist. The pipeline can also be utilized to detect these orbital regimes preferentially with a different choice in search criteria. All detected asteroid information is shown in the appendix in Table \ref{tab:night1} and Table \ref{tab:night2}.

\subsection{ETA Upper Limit}

\begin{figure}
  \includegraphics[width=\columnwidth]{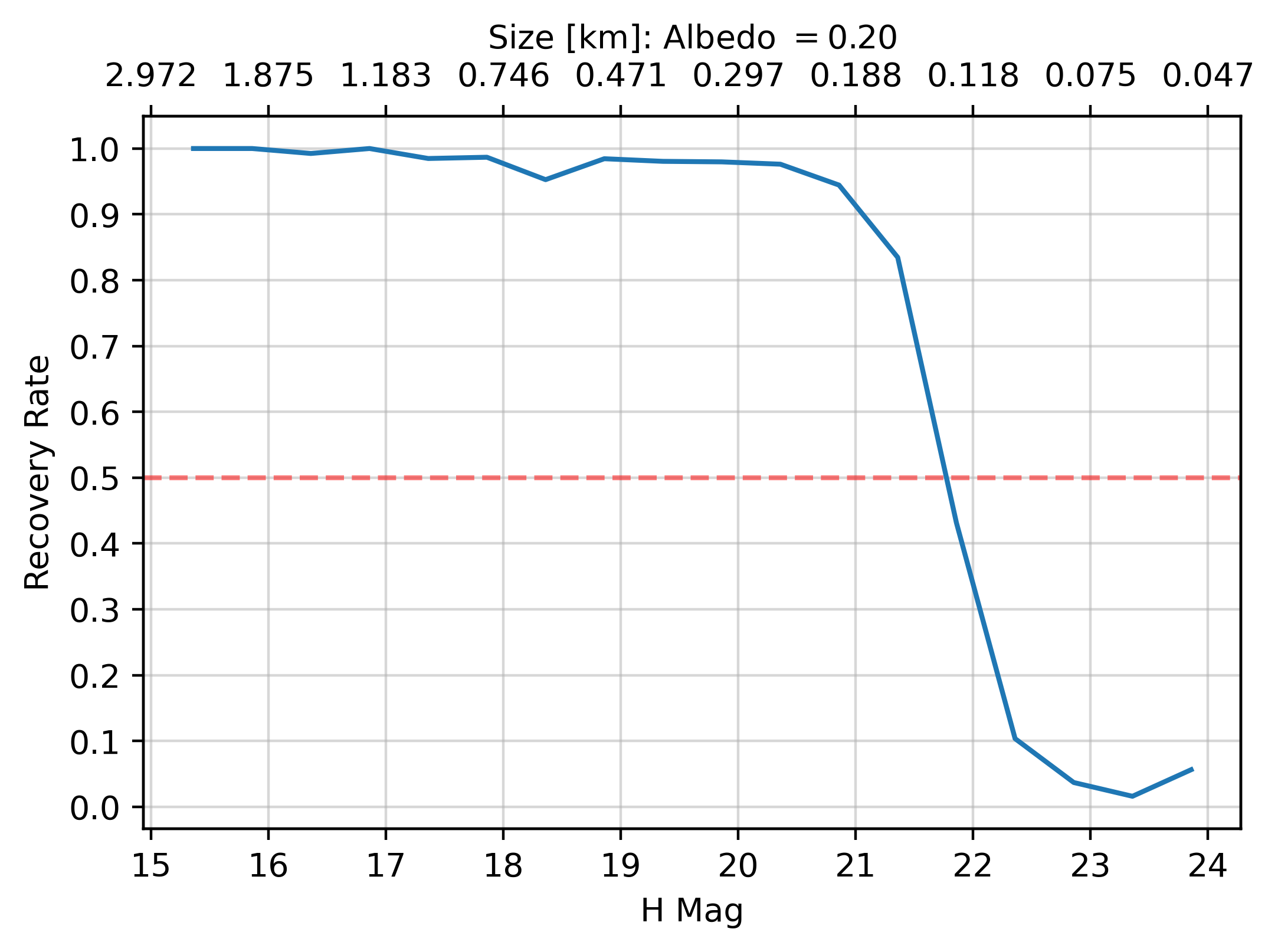}
  \caption{Recovery rate of injected asteroids as a function of absolute magnitude H. The red dotted line shows $50\%$ recovery and thus overall sensitivity, which was $21.77$. We convert to asteroid diameter in kilometers assuming the standard S-class asteroid albedo of 0.20.}
  \label{fig:recovery_rate_absolute}
\end{figure}

We convert from PAN-STARR's calibrated r-band magnitude to apparent magnitude V:
\begin{equation}
V = m_r + 0.255,
\end{equation} which assumes ETAs are most similar to S-class asteroids with albedo 0.2 \citep{2017arXiv170506209C}. We then convert to absolute magnitude H: \begin{equation}
H = V + 2.5\log_{10}(\phi), 
\end{equation} where the phase integral $\phi$ is given by
\begin{multline}
    \phi\left(G=0.15\right) = 0.85 \exp\left(-3.332\tan\left(\alpha/2\right)^{0.631}\right) + \\ 0.15\exp\left(-1.862 \tan\left(\alpha/2\right)^{1.218}\right),
\end{multline} and $\alpha$ is the Sun-asteroid-Earth angle \citep{2007JBAA..117..342D}. Here we assume a distance of 1 AU and angle of 60$\circ$ based on the L4 Lagrange point. Our recovery rate is $> 50\%$ for $V < 23.92$ and $H < 21.77$. Our recovery rate as a function of absolute magnitude H is shown in Figure \ref{fig:recovery_rate_absolute}. 

We calculate our simulated ETA population coverage using the time average position of stable ETAs over the course of the simulation shown in Figure \ref{fig:ETA_sim}. We find that our observations cover $0.625\%$ of the ETA population. There is some sensitivity to simulation initial conditions, but we find this distribution to be most representative of a potential ETA population given the large number of asteroids we injected into our simulation and the long timescale of our simulation during which the ETAs come into equilibrium. Our simulation is not as long as the theoretical lifetimes of long term stable ETAs ($2\times10^7$ compared to $10^8$ or $10^9$ years). It is the longest simulation of ETAs of its size, though, and provides one of the most accurate pictures of the ETA population distribution. While it is possible that certain population parameters can change in simulating from $2\times10^7$ to $10^8$ years, it is not likely that the time-averaged spatial distribution as viewed from earth would vary significantly. Our coverage is substantially lower than previous ETA searches, because we assume a much broader stable ETA population based on simulation results. Even with varying simulation initial conditions, our population is still substantially broader. We discuss coverage comparisons and the significance of our simulation in the constraint calculation in more detail in \S\ref{sec:comparison}.

We use a frequentist approach in calculating the upper limit on the ETA population following the single pipeline case of \citet{2009CQGra..26x5007S}. We have a Poisson distributed process for ETA detection with a probability of detecting N events:
\begin{equation}\label{eq:poisson}
  P(N|\epsilon \lambda ) = \frac{(\epsilon \lambda)^N}{N!}e^{-(\epsilon \lambda)},
\end{equation} where $\lambda$ is the expected mean number of events and $\epsilon$ is our pipelines probability of detecting a given event. The upper limit on expected number of events $\lambda$ can be calculated at a given confidence level $\alpha$ and given number of actually measured events $n$ by calculating the value of $\lambda$ at which there is a probability $\alpha$ of detecting more than $n$ events:
\begin{equation}\label{eq:poisson_confidence}
  \alpha = \sum_{N=n+1}^{\infty} P(N|\epsilon \lambda ) = 1 - \sum_{N=0}^{n} P(N|\epsilon \lambda ).
\end{equation} In our case, we have no detected events, $n=0$, and for a given absolute magnitude H, we have $\epsilon(H) = R(H)*C$, where $R(H)$ is our recovery as a function of absolute magnitude H and $C$ is our coverage percentage. We calculate an upper limit with 95$\%$ confidence:
\begin{equation}\label{eq:upper_limit}
    0.05 = P(0|\epsilon \lambda ) = e^{-(\epsilon \lambda)}, \lambda = \frac{3}{R(H)*C}.
\end{equation}

\begin{figure}
  \includegraphics[width=\columnwidth]{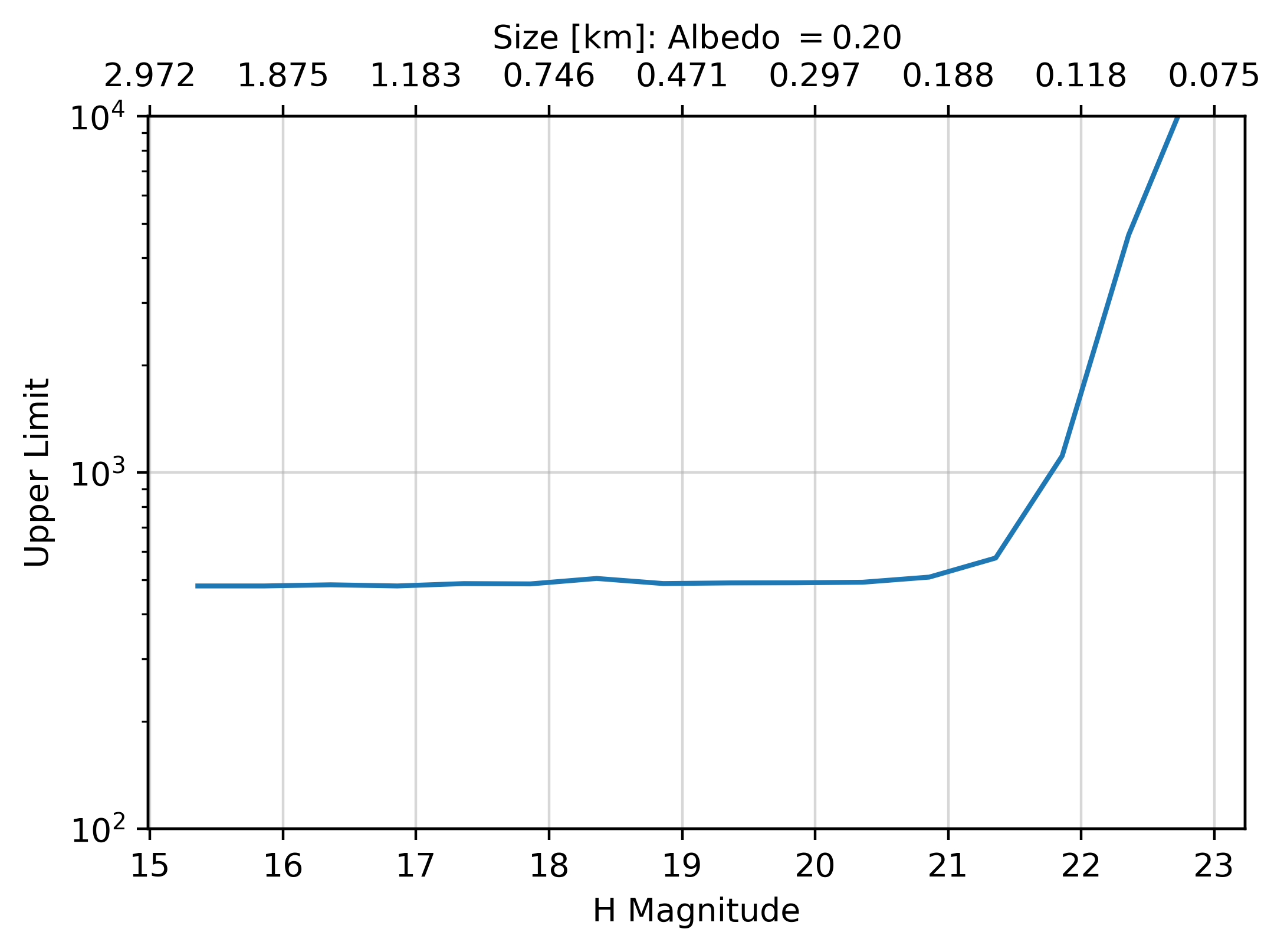}
  \caption{Upper limit of the ETA population at L4 directly calculated from our recovery rate. The flat area on the left is a result of maximum recovery with small coverage. The sloped area on the right is where our recovery starts to drop.}
  \label{fig:upper_limit_obs}
\end{figure}

 Our calculated upper limit is shown in Figure \ref{fig:upper_limit_obs}. This calculation is for the L4 Lagrange point only. We assume that L5 has the same upper limit by symmetry and so we also freely compare with previous searches of L5. The upper limit curve can be broken into two areas. The flat upper limit on the left and middle is a result of our maximum recovery rate through an H magnitude of $~20.5$. The calculated upper limit here is restricted by our coverage. The heavily sloped upper limit on the right is where our recovery rate drops below $50\%$ and becomes the greatest restriction in the upper limit calculation. Thus our strongest calculated upper limit is at the end of the flat area, where our recovery rate and limited coverage balance. We use results here to extrapolate to other H magnitudes to calculate a more accurate $U(H)$.

If we assume that ETAs follow a similar power law H-distribution as Near Earth Objects (NEOs), we can calculate a more stringent and accurate $U(H)$. We make use of the H distribution for NEOs:
\begin{equation}
  N(<H) = A*10^{\alpha H},
\end{equation}\label{eq:N} where $ \alpha = 0.48 \pm 0.02$ for $13<H<16$, $\alpha = 0.33 \pm 0.01$ for $16<H<22$, and $\alpha = 0.62 \pm 0.03$ for $H>22$ \citep{2017Icar..284..114S}. This is consistent with \citet{2015Icar..257..302H,2017Icar..284..416T,2005Natur.435..462M}. We take our strongest upper limit result of $N(<H=21.357)=62.84$ and extrapolate using Equation \ref{eq:N} (see Figure \ref{fig:upper_limit_broad_ours}).

\startlongtable
\begin{deluxetable*}{ccccc}
\tablenum{1}
\tablecaption{Constraint Analysis Results \label{tab:analysis}}
\tablewidth{0pt}
\tablehead{
\colhead{Study} & \colhead{Narrow ETA Population} & \colhead{Narrow ETA Population} & \colhead{Broad ETA Population} & \colhead{Broad ETA Population} \\ 
\colhead{\,} & \colhead{Markwardt H Distribution} & \colhead{Current H Distribution} & \colhead{Markwardt H Distribution} & \colhead{Current H Distribution}}
\startdata
 \text{This Study} & 8.74 & 22.40 & 80.03 & 205.05 \\
 \text{Markwardt} & 23.60 & 21.77 & 187.82 & 173.24 \\
 \text{Whiteley and Tholen} & 108.30 & 242.68 & 1740.81 & 3900.79 \\
 \text{Cambioni} & 40.45 & 49.93 &  &  \\
 \enddata
 \tablenotetext{}{Comparisons of ETA population constraints at a single Lagrange point at $H=20$ (297 meters at 0.2 albedo). We compare the constraints set by our analysis, \cite{2020MNRAS.492.6105M}, \cite{1998Icar..136..154W}, and \cite{2018LPI....49.1149C}.}
 \end{deluxetable*}

\section{Discussion} \label{sec:discussion} 

\subsection{Comparison to Previous Searches}\label{sec:comparison}

The two major differences between our upper limit calculation and previous searches are different H distributions and a broader ETA population used for determining coverage. The H distribution used in \citet{2018LPI....49.1149C,2020MNRAS.492.6105M} is a power law with $\alpha = 0.7$ and $\alpha = 0.46 - 0.7$, respectively. Our H distribution is a power law with $ \alpha = 0.48$ for $13<H<16$, $\alpha = 0.33$ for $16<H<22$, and $\alpha = 0.62$ for $H>22$ and is based on the most recent NEO population models. We have completed our analysis for the alpha values used by \citet{2020MNRAS.492.6105M} as well as our own for comparison. 

The previous upper limit calculations have used a very narrow ETA population distributions in determining observational coverage. We find that they use some cuts in determining a potential ETA population are overly strict, such as a restriction on heliocentric ecliptic longitude being within 30 degrees of the Lagrange point. Our simulation shows that long-term stable ETAs travel well beyond this cut. Additionally, previous calculations have not used simulations and instead have calculated the Jacobi integral for initial conditions only. While this integral is a valid representation of Trojan orbits, it is not as accurate as a long term simulation in determining time average positions of stable ETAs. We have completed our analysis for both the broad ETA population simulated in this paper as well as the more narrow population determined by \citet{2020MNRAS.492.6105M}.

We applied our analysis on our results as well as literature results \citep{1998Icar..136..154W, 2018LPI....49.1149C, 2020MNRAS.492.6105M}. We repeat our analysis for four situations: narrow ETA population with \citet{2020MNRAS.492.6105M} H distribution, narrow ETA population with our H distribution, broad ETA population with \citet{2020MNRAS.492.6105M} H distribution, and broad ETA population with our H distribution. We estimate our coverage on the narrow ETA population of \citet{2020MNRAS.492.6105M} by assuming that the variation in spatial ETA distribution over their observed area is minimal and multiplying our observed sky area by their coverage per square degree. We use the same instrument on the same telescope and point at nearly the exact same ecliptic location relative to respective Lagrange points. This gives us a coverage of $C=5.72$ for the narrow ETA population. We estimate the coverage of \citet{1998Icar..136..154W,2020MNRAS.492.6105M} of the broad ETA population using their given observational information. Unfortunately, \citet{2018LPI....49.1149C} provides too limited observational information, but their constraint is weaker than ours and that of \citet{2020MNRAS.492.6105M}. 

\begin{figure}
  \includegraphics[width=\columnwidth]{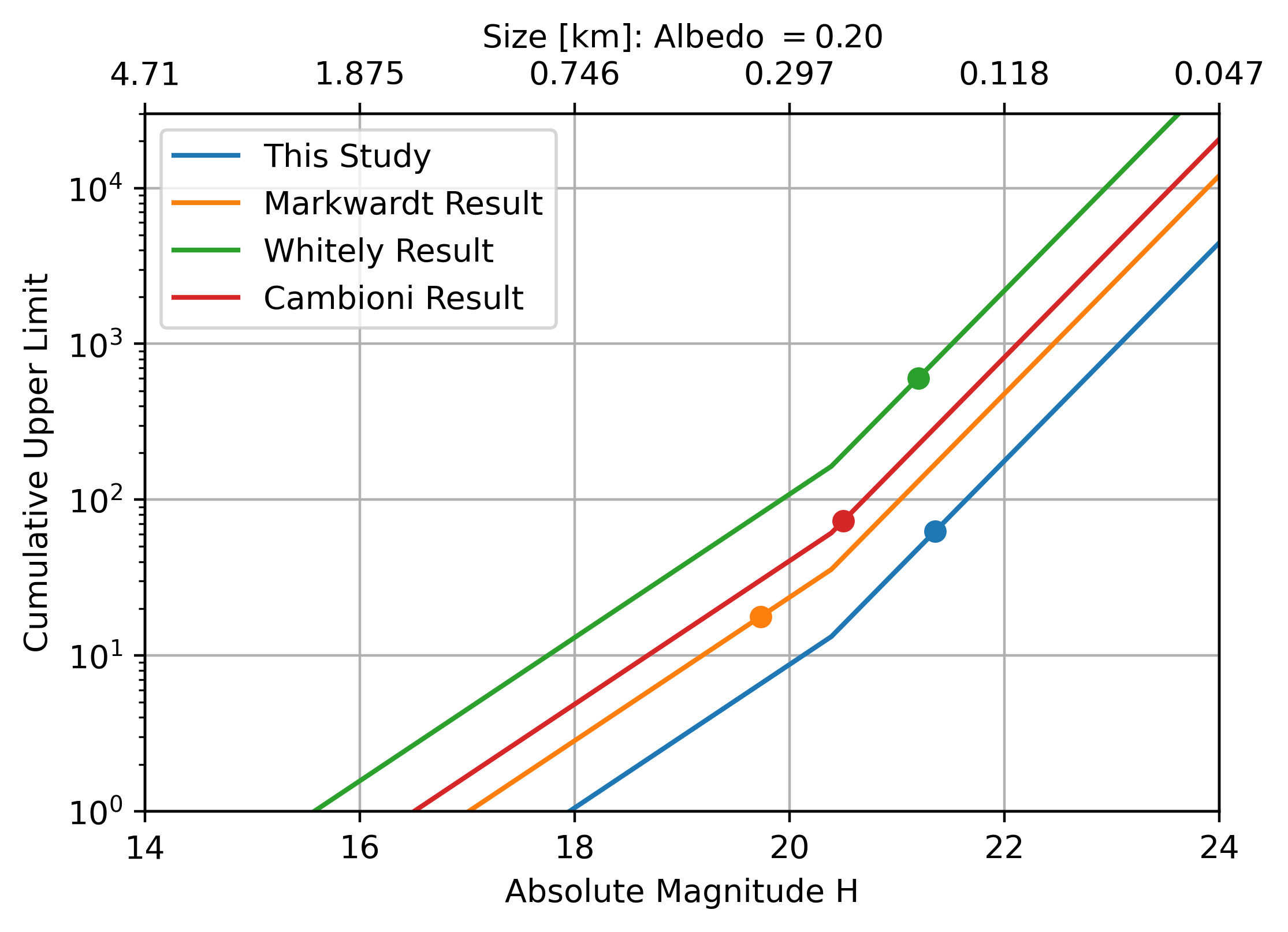}
  \caption{Upper limit of the ETA population at L4 using a narrow ETA population and the H-mag distribution of \citet{2020MNRAS.492.6105M}. Results are shown for this paper and the other three from the literature \citep{1998Icar..136..154W,2018LPI....49.1149C,2020MNRAS.492.6105M}.}
  \label{fig:upper_limit_narrow_mark}
\end{figure}

\begin{figure}
  \includegraphics[width=\columnwidth]{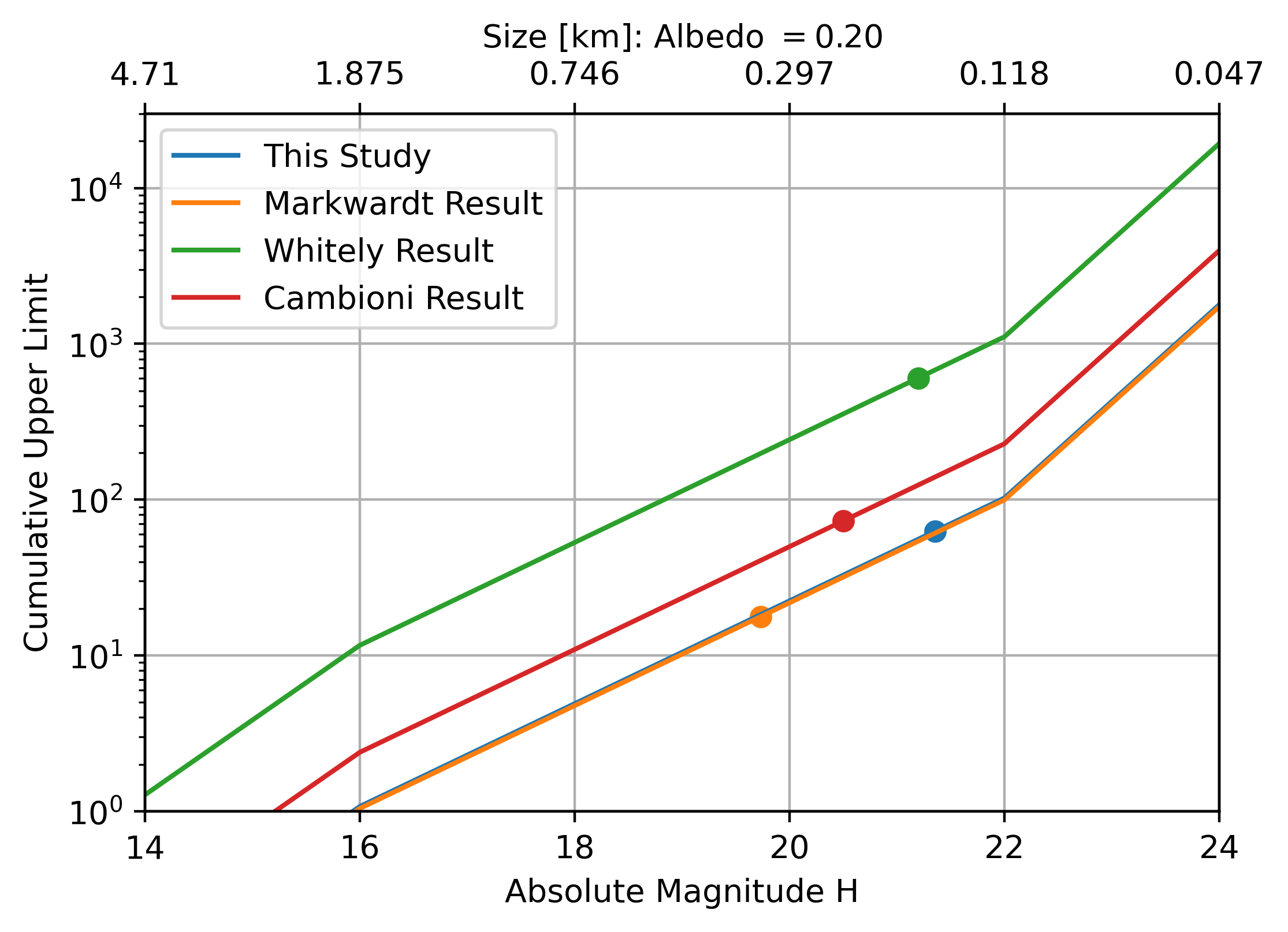}
  \caption{Upper limit of the ETA population at L4 using a narrow ETA population and our up-to-date H-distribution. Results are shown for this paper and the other three from the literature \citep{1998Icar..136..154W,2018LPI....49.1149C,2020MNRAS.492.6105M}.}
  \label{fig:upper_limit_narrow_ours}
\end{figure}

\begin{figure}
  \includegraphics[width=\columnwidth]{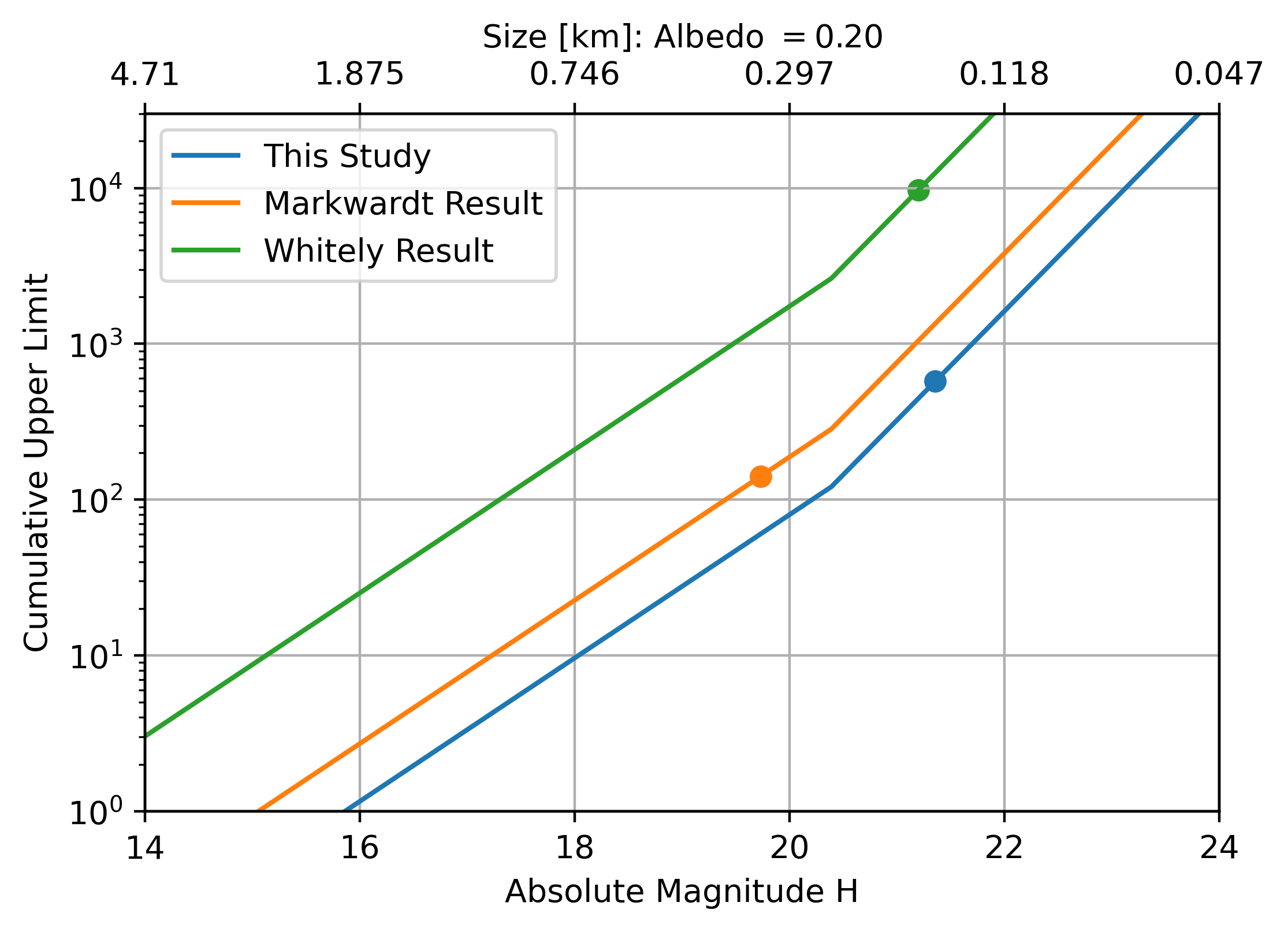}
  \caption{Upper limit of the ETA population at L4 using our broad ETA population and the H-mag distribution from \citet{2020MNRAS.492.6105M}. Results are shown for this paper along with \citet{1998Icar..136..154W,2020MNRAS.492.6105M}.}
  \label{fig:upper_limit_broad_mark}
\end{figure}

\begin{figure}
  \includegraphics[width=\columnwidth]{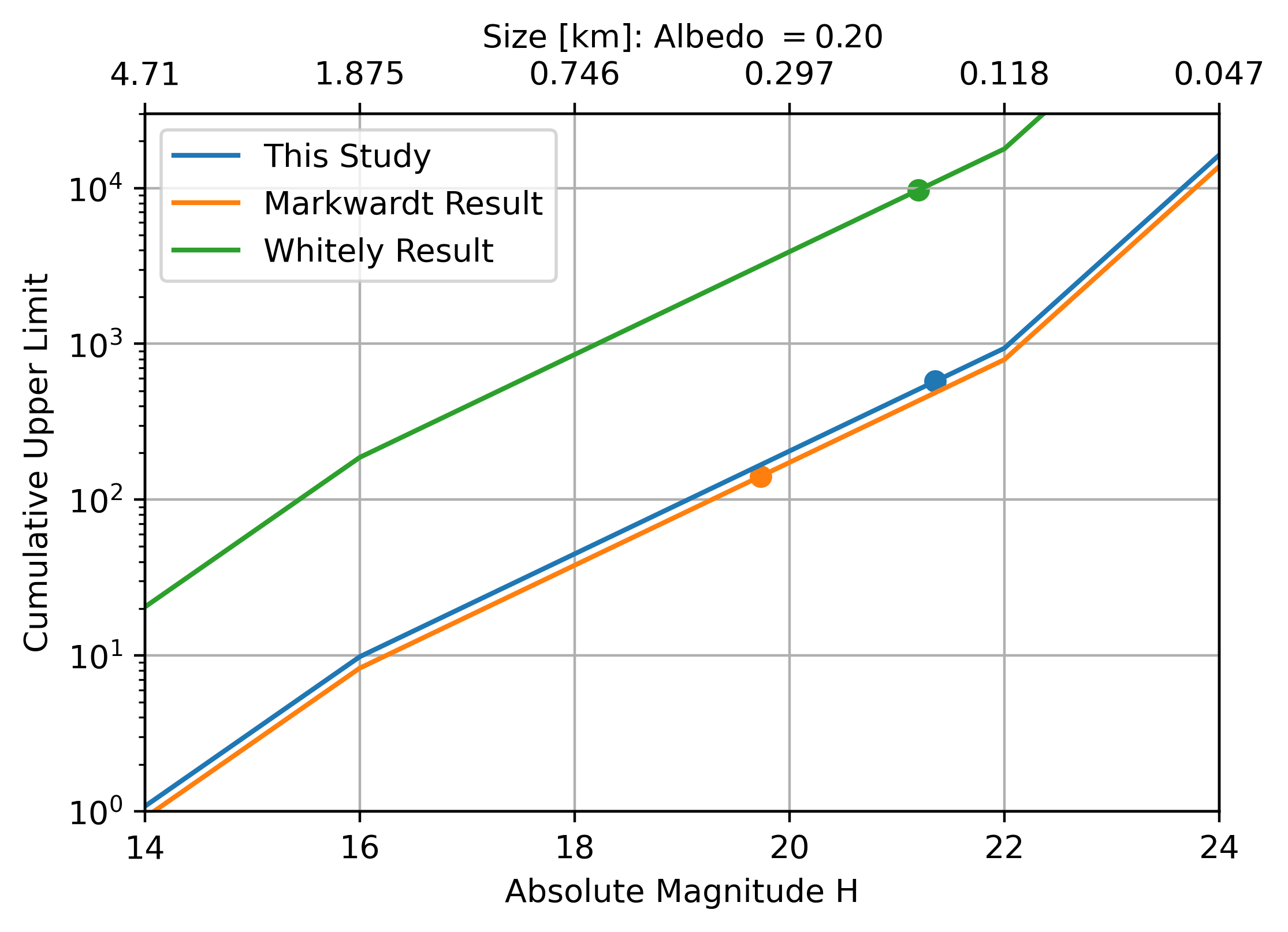}
  \caption{Upper limit of the ETA population at L4 using our broad ETA population and our up-to-date H-distribution. Results are shown for this paper, along with \citet{1998Icar..136..154W,2020MNRAS.492.6105M}. We consider this to be the most current and accurate representation of the ETA population. We find an L4 Earth Trojan Population of $N_{ET} < 1$ for $H = 13.93$, $N_{ET} < 10$ for $H = 16$, and $N_{ET} < 938$ for $H = 22$.}
  \label{fig:upper_limit_broad_ours}
\end{figure}

 We repeat our analysis for four situations, which are shown in Figures \ref{fig:upper_limit_narrow_mark}, \ref{fig:upper_limit_narrow_ours}, \ref{fig:upper_limit_broad_mark}, and \ref{fig:upper_limit_broad_ours} with results for $H=20$ shown in Table \ref{tab:analysis}. Here we assume the ETA population at L4 and L5 are identical, and we should note again that this analysis only applies to long-term stable ETAs and not captured ones. We consider \ref{fig:upper_limit_broad_ours} the most accurate and up to date constraint on the Earth Trojan Asteroid population. When using our more accurate H distribution, our upper limit is consistent with \citet{2020MNRAS.492.6105M}. When using the H distribution of \citet{2020MNRAS.492.6105M}, we have a stronger result. The broad ETA population significantly increases the upper limit calculated by all searches. The upper limit values depend heavily on assumed ETA population distribution, and could vary if a different population distribution is used. We are currently preparing a full analysis of different ETA simulations and potential population distributions (Yeager et al. in prep).

\subsection{Outlook for ETA Surveys}

In this paper we demonstrate that stacking signal across multiple epochs is an effective way to increase survey sensitivity for ETAs. This is not surprising, as this method has been used for many populations of asteroids. However, this technique has important limitations when applied to ETAs. All inner solar system asteroids enter the regime of non-linear motion within approximately a single night \citep[see Figure 4 of][for a clever demonstration of this]{2015AJ....150..125H}. This limits the time over which signal may be stacked to increase sensitivity. We have employed this fact to reach comparable depths of \citet{2020MNRAS.492.6105M} despite a quarter the on sky coverage. 

Where our method could excel beyond this limitation is in its track-before-detect framework, which enables detection even on non-linear trajectories. This requires the parameterization of the motion model in order to track for arbitrarily long timescales and requires the ability to quickly generate large numbers of trajectories. However, ETAs move under comparable influence of the Sun and the Earth, and thus their trajectories are not parameterizable by Keplerian orbital elements. There is promise in several machine learning methods (convolutional neural networks, generative adversarial network, and spatial statistical methods involving Gaussian processes) to develop surrogate models to quickly generate ETA trajectories to feed our track-before-detect pipeline. However, this is beyond the scope of this paper. We are eager to learn of new developments on this avenue as large scale twilight surveys could unlock incredible potential to detect ETAs. Until then, shift-and-stack methods on single night data sets carried out over increasingly more nights with larger telescopes and cameras with larger fields of view and shorter overheads will be the best we can do.

\acknowledgments

This work was performed under the auspices of the U.S. Department of Energy by Lawrence Livermore National Laboratory under Contract DE-AC52-07NA27344 and was supported by the LLNL-LDRD Program under Projects 17-ERD-120, 19-SI-004, and 20-ER-025. This work was based on observations obtained at Cerro Tololo Inter-American Observatory a division of the National Optical Astronomy Observatories, which is operated by the Association of Universities for Research in Astronomy, Inc. under cooperative agreement with the National Science Foundation.

\facilities{CTIO(Blanco/DECam)}

\appendix

\startlongtable
\begin{deluxetable*}{ccccccc}
\tablenum{2}
\tablecaption{Night One Detections \label{tab:night1}}
\tablewidth{0pt}
\tablehead{
\colhead{MPC Designation} & \colhead{V} & \colhead{R.A.} & \colhead{Dec.} & \colhead{$\pi_{\alpha}\cos{\delta}$\tablenotemark{a}} & \colhead{$\pi_{\delta}$} & \colhead{Observation Time \tablenotemark{b}} \\ 
\colhead{\,} & \colhead{Magnitude} & \colhead{$^\circ$} & \colhead{$^{\circ}$} & \colhead{$^\circ\,day^{-1}$} & \colhead{$^\circ\,day^{-1}$} & \colhead{MJD}}
\decimalcolnumbers
\startdata
 28445 & 19.2 & 299.00217 & -20.80504 & 0.4 & 0.046 & 58807.012465 \\
 6588 & 18.2 & 299.6919 & -20.43834 & 0.35 & 0.059 & 58807.012465 \\
 25990 & 19.0 & 300.15762 & -20.05189 & 0.569 & 0.082 & 58807.013264 \\
 18284 & 18.5 & 300.0047 & -20.69366 & 0.629 & 0.166 & 58807.012465 \\
 382363 & 22.0 & 300.01745 & -20.2974 & 0.626 & 0.119 & 58807.014861 \\
 218860 & 21.6 & 299.60271 & -19.96368 & 0.53 & 0.056 & 58807.012465 \\
 474354 & 22.2 & 298.72499 & -21.01137 & 0.565 & 0.143 & 58807.05162 \\
 71096 & 18.5 & 299.2128 & -20.24981 & 0.367 & 0.111 & 58807.012465 \\
 359369 & 22.4 & 300.60024 & -20.9554 & 0.643 & 0.106 & 58807.026817 \\
 428658 & 21.1 & 299.39684 & -20.28277 & 0.688 & 0.088 & 58807.012465 \\
 321997 & 21.3 & 299.72667 & -20.82516 & 0.634 & 0.089 & 58807.022037 \\
 66597 & 20.2 & 299.03489 & -20.69781 & 0.561 & 0.091 & 58807.012465 \\
 30031 & 19.5 & 299.52837 & -21.07535 & 0.493 & 0.139 & 58807.012465 \\
 2763 & 16.6 & 299.08874 & -19.91832 & 0.524 & 0.13 & 58807.022037 \\
 474354 & 22.2 & 298.72479 & -21.01091 & 0.651 & 0.139 & 58807.054028 \\
 158 & 15.0 & 299.31813 & -20.15988 & 0.313 & 0.061 & 58807.012465 \\
 933 & 18.0 & 300.0497 & -20.13273 & 0.319 & 0.068 & 58807.04206 \\
 200055 & 20.2 & 300.56029 & -20.24477 & 0.547 & 0.151 & 58807.012465 \\
 \enddata
 \tablenotetext{a}{$\alpha$ and $\delta$ refer to right ascension and declination, respectively.}
 \tablenotetext{b}{Observation time refers to the time at which the R.A. and Dec. values were observed. These were the first exposure the detected trajectory intersected.}
 \end{deluxetable*}
 
 \startlongtable
\begin{deluxetable*}{ccccccc}
\tablenum{3}
\tablecaption{Night Two Detections \label{tab:night2}}
\tablewidth{0pt}
\tablehead{
\colhead{MPC Designation} & \colhead{V} & \colhead{R.A.} & \colhead{Dec.} & \colhead{$\pi_{\alpha}\cos{\delta}$} & \colhead{$\pi_{\delta}$} & \colhead{Observation Time} \\ 
\colhead{\,} & \colhead{Magnitude} & \colhead{$^\circ$} & \colhead{$^{\circ}$} & \colhead{$\circ / day$} & \colhead{$\circ / day$} & \colhead{MJD}}
\decimalcolnumbers
\startdata
& 19.7 & 300.81914 & -20.28069 & 0.559 & 0.171 & 58808.012176 \\
 6429 & 18.3 & 299.92129 & -20.854 & 0.579 & 0.097 & 58808.012176 \\
 19890 & 20.5 & 300.67434 & -20.1136 & 0.41 & 0.184 & 58808.012176 \\
 79637 & 20.2 & 300.10334 & -19.72785 & 0.499 & 0.076 & 58808.014583 \\
 933 & 18.0 & 300.39079 & -20.0857 & 0.329 & 0.053 & 58808.012176 \\
 6186 & 18.6 & 301.1003 & -19.93354 & 0.404 & 0.102 & 58808.053611 \\
 7316 & 19.9 & 301.4632 & -21.00672 & 0.305 & 0.119 & 58808.012176 \\
 30031 & 19.5 & 300.06307 & -20.93398 & 0.508 & 0.153 & 58808.012176 \\
 137296 & 20.4 & 301.31721 & -20.52184 & 0.56 & 0.089 & 58808.012176 \\
 33639 & 18.7 & 300.41655 & -20.58602 & 0.346 & 0.014 & 58808.012176 \\
 11464 & 20.9 & 299.81264 & -20.09294 & 0.241 & 0.056 & 58808.012176 \\
 428658 & 21.1 & 300.13527 & -20.19834 & 0.687 & 0.089 & 58808.012176 \\
 34004 & 18.5 & 301.62358 & -20.22192 & 0.532 & 0.123 & 58808.012176 \\
 33056 & 20.3 & 299.76062 & -20.77199 & 0.359 & 0.113 & 58808.064757 \\
 32002 & 19.7 & 300.33409 & -20.1858 & 0.517 & 0.08 & 58808.012975 \\
 153340 & 20.0 & 301.44698 & -20.4279 & 0.559 & 0.154 & 58808.032176 \\
 4795 & 18.3 & 300.93586 & -21.21965 & 0.514 & 0.08 & 58808.012176 \\
 183248 & 21.3 & 300.20088 & -20.77253 & 0.493 & 0.069 & 58808.012176 \\
 38240 & 20.2 & 300.57507 & -20.75502 & 0.5 & 0.08 & 58808.012176 \\
 67552 & 20.7 & 300.92415 & -19.76929 & 0.597 & 0.091 & 58808.036921 \\
 27162 & 18.7 & 300.09441 & -20.9108 & 0.389 & 0.112 & 58808.013785 \\
 245878 & 21.1 & 300.24139 & -19.76707 & 0.485 & 0.086 & 58808.036921 \\
 44031 & 20.8 & 300.9262 & -19.56985 & 0.376 & 0.034 & 58808.019433 \\
 67729 & 20.6 & 301.67756 & -20.77597 & 0.627 & 0.1 & 58808.012975 \\
 22520 & 19.4 & 300.55075 & -20.0181 & 0.558 & 0.098 & 58808.012176 \\
 40655 & 19.8 & 300.06836 & -20.56463 & 0.508 & 0.081 & 58808.012176 \\
 6588 & 18.2 & 300.06775 & -20.37796 & 0.342 & 0.08 & 58808.012176 \\
 18284 & 18.5 & 300.68213 & -20.5291 & 0.63 & 0.165 & 58808.012176 \\
 98050 & 20.3 & 301.10895 & -20.68522 & 0.548 & 0.132 & 58808.012176 \\
 6421 & 18.4 & 301.10738 & -20.32112 & 0.38 & 0.065 & 58808.012176 \\
 138278 & 20.3 & 301.56737 & -20.94839 & 0.475 & 0.111 & 58808.012176 \\
 200055 & 20.2 & 301.14832 & -20.0934 & 0.556 & 0.145 & 58808.012176 \\
 46863 & 20.1 & 301.28386 & -20.2624 & 0.349 & 0.073 & 58808.052824 \\
 37461 & 20.8 & 300.73613 & -19.76455 & 0.556 & 0.099 & 58808.024236 \\
 124069 & 21.1 & 301.09555 & -21.41326 & 0.246 & 0.13 & 58808.056019 \\
 62052 & 20.4 & 300.06279 & -20.11545 & 0.37 & 0.088 & 58808.012176 \\
 322045 & 21.7 & 300.09013 & -21.20186 & 0.583 & 0.129 & 58808.05838 \\
 \enddata
 \end{deluxetable*}

\bibliography{ETA}{}
\bibliographystyle{aasjournal}

\end{document}